\documentstyle[preprint,aps,epsf,floats]{revtex}
\tighten
\begin{document}

\preprint{\vbox{\hbox{JHU--TIPAC--96004}
\hbox{hep-ph/9602410}}}

\title{Scale Setting in Top Quark Decays}
\author{Thomas~Mehen}
\address{Department of Physics and Astronomy,
The Johns Hopkins University\\
3400 North Charles Street,
Baltimore, Maryland 21218 U.S.A.\\
{\tt mehen@dirac.pha.jhu.edu}}

\date{Apr 1996}

\maketitle
\begin{abstract}
We compute the $O(\beta_0 \alpha_s^2)$ QCD corrections to the Standard Model decay $t\rightarrow W^+ b$ as well as the non standard decay $t\rightarrow H^+ b$. We then use our results to compute the BLM scale for these decays, and study the dependence of the BLM scale on the mass of the boson in the decay. We find that the BLM procedure gives extremely small scales when top decays into heavy bosons. When we try to improve the BLM scale by reexpressing rates as a functions of the top quark running mass, we find the BLM scale exhibits unnatural dependence on the boson mass, casting doubts on the applicability of the BLM procedure to these processes. 

\end{abstract}

\pagebreak
 
The top quark, recently observed by CDF~\cite{CDF} and D0~\cite{D0},
will be the focus of much experimental attention in coming years. It will be important to test the Standard Model predictions for the width of the top quark as well as search for rare or non-standard decays which may give us a glimpse into post-Standard Model physics. Because of this it is important to have 
precise theoretical predictions for top quark decay, both for the
Standard Model and for theories which propose to go beyond it. 

In this paper, we calculate the $O(\beta_0 \alpha_s^2)$ corrections to the decays $t\rightarrow W^+ b$ and $t\rightarrow H^+ b$, including the full dependence on the mass of the boson in the decay. (We will neglect the mass of the b quark, however.) In many QCD calculations, these corrections form the dominant part of the complete $O(\alpha_s^2)$ correction. As examples we present the perturbative corrections to $R(e^+ e^- \rightarrow {\rm hadrons})$, $\Gamma(\tau \rightarrow \nu_{\tau}~{\rm hadrons})$ and the series relating the pole mass and the $\overline{MS}$ mass of a heavy quark:

\begin{equation}\label{mass}
m^{pole}_Q = m^{\overline{MS}}_Q\left[1 + {4 \alpha(m_Q) \over 3 \pi} +
           (1.56 \beta_0 - 1.05) {\alpha(m_Q)^2 \over \pi^2}\right],
\end{equation}

\begin{equation}
R(e^+ e^- \rightarrow {\rm hadrons}) =
3 \left( \sum_i Q_i^2 \right) \left[1 + {\alpha(\sqrt{s}) \over \pi} +
           (0.17 \beta_0 + 0.08) {\alpha(\sqrt{s})^2 \over \pi^2}\right],
\end{equation}

\begin{equation}
\Gamma(\tau \rightarrow \nu_{\tau}~{\rm hadrons}) = 3~\Gamma(\tau \rightarrow \nu_{\tau} \overline{\nu_e} e^-) \left[1 + {\alpha(m_{\tau}) \over \pi} + (0.57 \beta_0 + 0.08) {\alpha(m_{\tau})^2 \over \pi^2}\right].
\end{equation}
In each case the $O(\beta_0 \alpha_s^2)$ term is a good approximation to the complete $O(\alpha_s^2)$ correction. Because of this, the calculation of the $O(\beta_0 \alpha_s^2)$ corrections for top quark decays is of interest in its own right. This calculation is a principal result of this paper.

The $O(\beta_0 \alpha_s^2)$ corrections can also be used to set the scale of the QCD coupling, via the scheme of Brodsky, Lepage, and Mackenzie~\cite{BLM}. Varying the scale of the QCD coupling changes the $n^{th}$ order coefficient in the perturbation series by a term proportional to $\beta_0^{n-1} \alpha_s^n$. Since $\beta_0$ is numerically rather large, it is clear that a poor choice in scale can result in a poorly behaved perturbative expansion. The BLM scheme sets the scale so that the $O(\beta_0 \alpha_s^2)$ term is cancelled. 

The BLM scale is commonly used as a criterion for assessing the validity of a perturbative QCD series. A small BLM scale is often interpreted as a sign that higher order corrections are not well behaved and that the perturbation series is not to be trusted.  However, for the top quark decays discussed in this paper, we will see that the BLM scale can become much smaller than any of the relevant mass scales in the problem. For instance, we will find a BLM scale of 200 MeV for a 175 GeV top quark decaying into 150 GeV Higgs and massless b quark. Since the energy released in the decay is far greater than the QCD scale, it is hard to believe that perturbative QCD is really in trouble. Indeed, by examining the coefficients in the perturbative series we will see that the small BLM scale is a consequence of a small leading order coefficient rather than a large second order coefficient.  For top quark decaying into very heavy Higgs, the BLM scale ceases to be a useful tool in analyzing the validity of the perturbative QCD series. 

More sophisticated attempts at getting at the ``true'' scale of a leading order calculation were made in refs.~\cite{REN,BB}. These authors resum renormalon chains, which include the entire $\beta_0^{n - 1} \alpha_s^n$ series, for processes such as ${\rm R}({\rm e}^+{\rm e}^- \rightarrow {\rm hadrons})$, $\Gamma (\tau \rightarrow \nu_{\tau}~{\rm hadrons})$, and $\Gamma({\rm b}\rightarrow {\rm c}{\rm e}\nu)$. This can be viewed as an all orders generalization of the BLM scheme, which only sums the $O(\beta_0 \alpha_s^2)$ term correctly. 
The sum of this infinite series, defined as the principal value of the Borel integral, can be absorbed into the scale of the coupling constant in the leading order calculation. The efffective scale found in these calculations rarely differs from the BLM scale by more than $\pm 50\%$, indicating that the BLM scale may be used to obtain a rough estimate of the sum of this subclass of higher order corrections. The quality of the perturbative series is not measured by this effective scale, however, but by the imaginary part of the Borel integral, which is interpreted as being the intrinsic uncertainty which one must encounter when trying to sum a series which is not Borel summable. 
In ref. \cite{BB}, a renormalon summation is done for the decay $t \rightarrow W^+ b$ in the limit $M_W \rightarrow 0$. The authors of \cite{BB} find an effective scale of $0.89~m_t$, to be compared to the BLM scale $0.122~m_t$ \cite{SV}. The uncertainty in the summation is estimated to be less than one percent, in line with our intuition that strong interaction corrections to this decay should be under perturbative control.  It will be of interest to redo the calculation allowing for finite W and Higgs masses. Such a calculation is currently in progress \cite{Mehen}.

We begin by considering the Standard Model top quark decay $t\rightarrow b W^+$. The QCD corrections to the tree level decay rate have been computed to $O(\alpha_s)$ in ref. \cite{JK}:
\begin{equation}\label{tWb}
 \Gamma(t\rightarrow W^+ b) = \Gamma_0\left[1 + {2 \alpha_s(\mu)\over 3 \pi} 
 \left( - {2 \pi^2 \over 3} + {5 \over 2} + \delta\left(x_W\right) \right)\right],
\end{equation}
where $x_W = m_W^2/m_t^2$, and $\Gamma_0$, the tree level decay rate, is 
\begin{equation}\label{twb0}
   \Gamma_0 = |V_{tb}|^2 {G_F m_t^3 \over 8{\sqrt 2} \pi}
       \left( 1 + 2 x_W\right)
       \left( 1 - x_W\right)^2
       = 1.33\:{\rm GeV}.
\end{equation}
In eqns.~(\ref{tWb}) and (\ref{twb0}), we have neglected the mass of the b-quark. We use $m_t = 175~{\rm GeV}$, $m_W = 80~{\rm GeV}$, and $V_{tb} = 1$. For these values of $m_W$ and $m_t$, $x_W =0.2$; $\delta(0.2) = 0.31$.

Smith and Voloshin~\cite{SV} have derived the following integral representation for the $O(\beta_0 \alpha_s^2)$ term in top quark decay:
\begin{equation}\label{SVIntegral}
   \delta \Gamma^{(2)} = - {\beta_0 \alpha_s^{(V)}(m_t) \over 4 \pi} 
         \int _0^{\infty}
         \left( \Gamma^{(1)}(\mu)-{m_t^2 \over \mu^2 + m_t^2}\Gamma^{(1)}(0)    
                  \right)   
          {d \mu^2 \over \mu^2}.
\end{equation}
Here, $\Gamma^{(1)}(\mu)$ is the one loop correction computed as if the 
gluon had mass $\mu$ and $\alpha_s^{(V)}(m_t)$ is the strong coupling constant defined in the V-scheme of Brodsky, Lepage, and MacKenzie~\cite{BLM}. The V-scheme coupling constant is related to the $\overline{MS}$ coupling constant by 
\begin{equation}
   \alpha_s^{(V)}(\mu)  = \alpha_s^{(\overline{MS})}(e^{-{5\over 6}}\mu)(1 + O(\alpha_s)).
\end{equation}
Computing in the limit of $m_W \rightarrow 0$, Smith and Voloshin find 
\begin{equation}
\delta \Gamma^{(2)} = - 1.73 \beta_0\left( {\alpha_s^{(V)}(m_t) \over \pi} \right)^2.
\end{equation}
Their result is confirmed in ref.~\cite{AC2}, where the calculation is done by direct evaluation of the $n_f$ dependent $O(\alpha_s^2)$ diagrams. Smith and Voloshin use this result to find a BLM scale of $0.122\: m_t$ (in the $\overline{MS}$ scheme) for top quark decay.  

\begin{figure}
\epsfxsize=9cm
\hfil\epsfbox{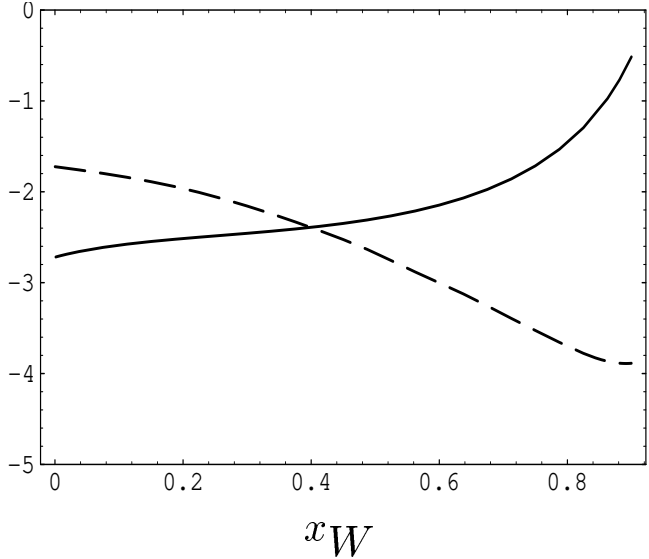}\hfill
\caption{Coefficients in perturbative expansion for $t \rightarrow W^+ b$;  $f(x_W)$ - solid, $g(x_W)$ - dashed.} 
\label{tbwCoeff}
\end{figure}

In this paper, we redo the calculation of $O(\beta_0 \alpha_s^2)$ corrections to $t \rightarrow W^+ b$ including the effect of finite W boson mass. We calculate the $O(\alpha_s)$ corrections to $t \rightarrow W^+b$, giving the gluon a finite mass. Integrals over Feynman parameters in the virtual gluon correction and over phase space in the brehmstrahlung graphs are computed numerically. The infrared divergences in each of these calculations cancel in their sum. Taking the gluon mass to zero we find our numerical result agrees with the analytic expressions in ref. \cite{JK}. We then use eq.~(\ref {SVIntegral}) to compute the $O(\beta_0 \alpha_s^2)$ corrections to $t \rightarrow W^+ b$. The rate is given by
\begin{equation}
 \Gamma(t\rightarrow W^+ b) = \Gamma_0\left[1 + f(x_W){\alpha_s(m_t) \over \pi} + g(x_W) \beta_0 {\alpha_s^2(m_t) \over \pi^2} \right],
\end{equation}
where the functions $f(x_W)$ and $g(x_W)$ are shown in Fig.~\ref{tbwCoeff}. The analytic expression for $f(x_W)$ can be found in ref. \cite{JK}.  For $x_W =0.2$, $f(x_W)=-2.51,~g(x_W)=-1.96$.

\begin{figure}
\epsfxsize=10cm
\hfil\epsfbox{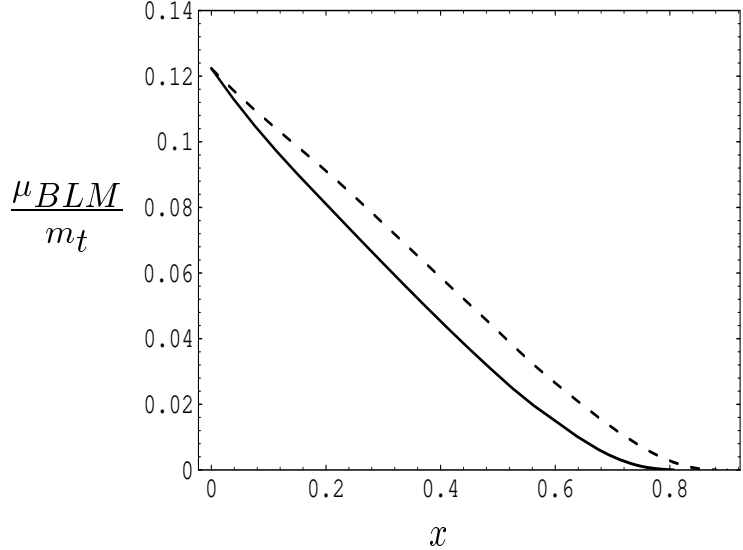}\hfill
\caption{BLM scale for top quark decays. The dotted line is for the decay $t\rightarrow W^+b$; $x = m_W^2/m_t^2$. For $m_W^2/m_t^2 = 0.2$, 
$\mu_{BLM} = 0.09\: m_t$. The solid line is for $t\rightarrow H^+b$; $x = m_H^2/m_t^2$.}
\label{BLM}
\end{figure}

The BLM scale, given by 
\begin{equation}\label{pole}
\mu_{BLM}(x_H) = m_t~{\rm exp}\left[-{2g(x_H)\over f(x_H)}\right].,
\end{equation}
is  plotted in Fig.~\ref{BLM} as a function of $x_W$. For $x_W = 0.2$, we find $\mu_{BLM} = 0.09~m_t$. If we use this scale for $\alpha_s $ in eq.~(\ref{tWb}), we find that QCD corrections reduce the width of the top quark by approximately 13\%. (For $\alpha_s$ we use the one-loop expression for the running coupling constant in the $\overline{MS}$ scheme, setting $n_f = 5$ and $\Lambda_{\overline{MS}} = 110\: {\rm MeV}$.) If we choose to leave the scale for the coupling equal to $m_t$ and include the $O(\beta_0 \alpha_s^2)$ term, the top quark decay rate is reduced by 11\%.  The effect of BLM scale setting on the theoretical prediction is small because for small $x_W$, $\alpha_s(\mu_{BLM})/\pi$ is still perturbative.  

Our computation of the BLM scale agrees with a similar calculation in ref.~\cite{LSW}, which studies the perturbative QCD corrections to the semileptonic decay of the bottom quark. 
These authors compute the BLM scale for the differential rate $d \Gamma /dq^2$, where $q^2$ is the invariant mass squared of the lepton pair.
Since the QCD corrections to the semileptonic decay are identical to those for the decay of top into a b quark and real W boson, we can simply substitute $m_t$ for $m_b$ and $m_W^2$ for $q^2$ and recover their result from ours. 

\begin{figure}
\epsfxsize=10cm
\hfil\epsfbox{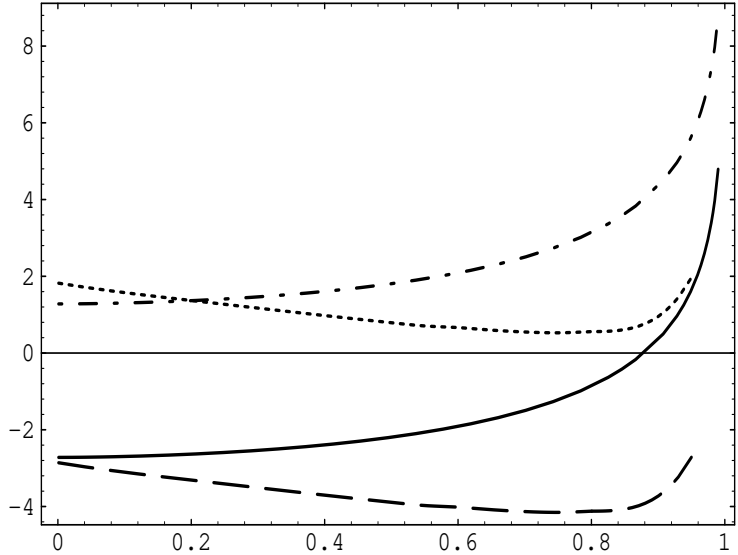}\hfill
\caption{Coefficients in perturbative expansions for $t \rightarrow H^+ b$. $f(x_H)$ - solid; $g(x_H)$ - dashed; $f(x_H)+4$ - dashed-dotted; $g(x_H) + 4.68$ - dotted.} 
\label{F&G}
\end{figure}

Next we would like to consider the QCD corrections to the decay of the top quark into a b quark and charged Higgs boson. A charged Higgs boson appears in any extension of the Standard Model which has more than one Higgs doublet. In general, it can have scalar or pseudoscalar couplings, so we parametrize the Feynman rule for the $t-H^+-b$ vertex as $i \overline{u}(b)(s + p \gamma_5)u(t)$, where the parameters $s$, $p$ depend on the specific model for the Higgs sector. The tree level decay rate (in the limit $m_b \rightarrow 0$) is

\begin{equation}\label{tHb_tree}
\Gamma_0(t \rightarrow H^+ b) = {m_t \over 16 \pi} 
    \left[ (|s|^2 + |p|^2) \left(1 - {m_H^2 \over m_t^2}
    \right)^2 \right].
\end{equation}
The $O(\alpha_s)$ corrections to this decay were calculated in ref.~\cite{LY}. 
\begin{equation}\label{tHb_oneloop} \nonumber
\Gamma(t \rightarrow H^+ b) =  \Gamma_0
       \left[1 + {2 \alpha_s(\mu)\over 3 \pi}\left(
     -{2 x_H \over 1 - x_H}{\rm ln}(x_H) + {2 \over x_H}{\rm ln}(1 - x_H)  \right.\right.
\end{equation}
\begin{eqnarray} 
\left. \left.- 5\:{\rm ln}(1 - x_H)- 2\:{\rm Li}_2(x_H)+2\:{\rm Li}_2(1-x_H) +{9 \over 2}-\pi^2 \right)\right], \nonumber
\end{eqnarray}
where $x_H = m_H^2/m_t^2$. The procedure for computing the $O(\beta_0 \alpha_s^2)$ corrections to $t \rightarrow H^+ b$ is identical to that for $t \rightarrow W^+ b$ discussed earlier. Our numerical calculations for the one loop corrections agree with eq.~(\ref {tHb_oneloop}) when we take the gluon mass to zero.  For $t \rightarrow H^+ b$ decay, we define coeffiecient functions $f(x_H)$ and $g(x_H)$ exactly as we defined $f(x_W)$ and $g(x_W)$ for $t \rightarrow W^+ b$. $f(x_H)$ and $g(x_H)$ are plotted in Fig.~\ref{F&G}. (The logarithmic divergence of these functions as $x_H \rightarrow 1$ is an artifact of the approximation $m_b = 0$.) 

The BLM scale for $t\rightarrow H^+ b$ is plotted in Fig.~\ref{BLM} as a function of $x_H$. Because the energy available for the gluon decreases as the mass of the boson increases, we expect the relevant scale for the QCD coupling to decrease as $m_W,m_H \rightarrow m_t$. What is surprising is that for both decays the BLM scale effectively vanishes well before the kinematic limit. For example, if we were to use the criterion $\alpha_s(\mu_{BLM})/ \pi \leq 1$ for a well behaved  perturbation series, we would conclude that the  perturbative computation of QCD corrections to top decay into charged Higgs becomes unreliable for $x_H \geq 0.75$ or $m_H \geq 150~{\rm GeV}$, $25 {\rm GeV}$ below the kinematic limit $m_t - m_b \simeq 175~{\rm GeV}$.
  
Upon examining Fig.~\ref{F&G}, it is clear that the small BLM scale obtained as $x_H \rightarrow 1$ is as much a consequence of the first order correction becoming small as it is a consequence of second order correction becoming large. It is also useful to look at the the numerical values of each term in the series. Considering the case $x_H = 0.75$ we find  
\begin {equation}
\Gamma = \Gamma_0 \left[1 - 1.22{\alpha_s(m_t) \over \pi} -
31.8 {\alpha_s(m_t)^2 \over \pi^2} \right] = \Gamma_0\left[1 - 0.043 - 0.039 \right].
\end{equation}
The second order correction is almost as large as the first; this is why we get the very low BLM scale 200 MeV. Using this BLM scale in the leading order expression we find the QCD corrections to the rate are $ -56\%$. On the other hand, the absolute size of each correction is small, and the expansion parameter $\alpha(m_t)/ \pi$ is only $\simeq 0.003$. To get a mere 4\% correction to the total rate would require a third order term of the size $\simeq 900~\alpha^3/\pi^3$ or $\simeq 15~\beta_0^2 \alpha^3/\pi^3$. This is not unreasonable, but it does require a large coefficient in the third order term. To get the kind of correction implied by the BLM method implies huge coefficients at higher orders. It seems likely that in this case the BLM scale setting method is overestimating the effect of higher order corrections.

Next, we will attempt to improve the BLM scale by rewriting the rate as a function of the top quark running mass. It is shown in ref. \cite{BBZ} that the leading renormalon singularity in the perturbative series for heavy quark decay is cancelled when the rate is expressed in terms of the running mass instead of the pole mass. Since the asymptotic growth of coefficients in the perturbative series is controlled by  renormalon singularites, one expects smaller coefficients in the perturbative series if we rewrite the series for top quark decay in terms of running top quark mass instead of the pole mass. While it is not obvious that arguments based on the asymptotic behavior of the perturbative series are relevant to the lowest order terms, it is in fact true that the coefficients of low order $O(\beta_0^n \alpha_s^{n+1})$ terms in the decay $t \rightarrow W^+ b$ (in the limit $M_W \rightarrow 0$) are significantly reduced when the series is expressed as a function of the running mass \cite{BB}. 

\begin{figure}
\epsfxsize=11cm
\hfil\epsfbox{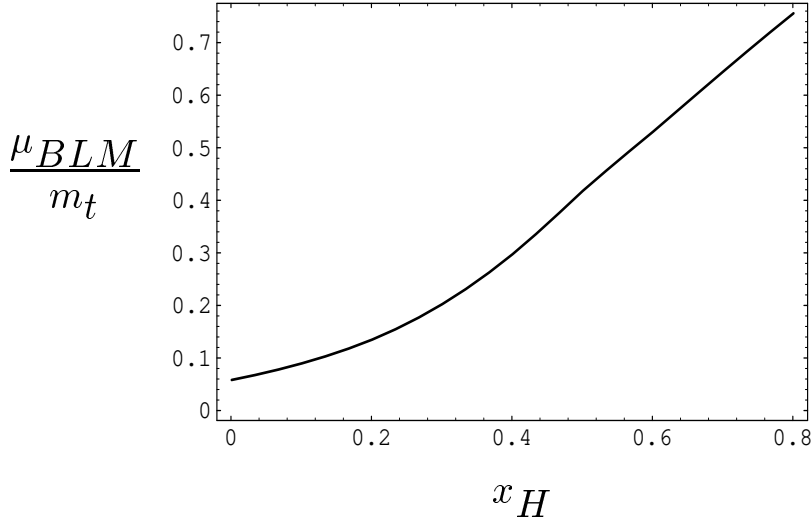}\hfill
\caption{BLM scale for $t \rightarrow H^+ b$ when rate is expressed as a function of the top quark $\overline{MS}$ mass instead of the pole mass.} 
\label{BLM2}
\end{figure}

In the limit $m_b \rightarrow 0$, the parameters $s$, $p$ are proportional to $m_t$, so the total rate is proportional to $m_t^3$.  Making the substitution in eq.~(\ref{mass}), then applying the BLM procedure to the new perturbative series results in the BLM scale shown in Fig.~\ref{BLM2}. The BLM scales computed for the two series are drastically different. The series obtained when the rate is expressed as a function of the $\overline{MS}$ mass has a higher BLM scale for almost all values of $x_H$. However, the dependence of the new BLM scale on $x_H$ seems unphysical. The BLM scale shown in Fig.~\ref{BLM2} actually becomes larger as $x_H \rightarrow 1$. 

To understand why the BLM scale changes so dramatically, we examine the coefficients of the one and two loop corrections for the new series.
When expressed as a function of the running top quark mass, the top quark width is given by:
\begin{equation}\label{msbar}
\Gamma \sim \overline{m_t}^3(m_t) \left[1 + (f(x_H) +4){\alpha_s(m_t) 
\over \pi} + (g(x_H)+ 4.68) \beta_0 {\alpha_s^2(m_t) \over \pi^2} \right],
\end{equation}
with a BLM scale
\begin{eqnarray}
\mu_{BLM} = m_t~{\rm exp}\left[-{2g(x_H)+9.36 \over f(x_H) + 4} \right]\nonumber
\end{eqnarray}
The functions $f(x_H)+4$ and $g(x_H)+4.68$ are also plotted in Fig.~\ref{F&G}.
In this case the ratio $(g(x_H)+4.68)/(f(x_H)+4)$ is small, especially for large $x_H$, and we find much larger BLM scales. As $x_H \rightarrow 1$, the replacement of the pole mass with the running mass leads to a nearly exact cancellation of the two loop correction. Since there is no physical reason to expect smaller higher order corrections in the limit $x_H \rightarrow 1$ it is doubtful that we should regard this cancellation as anything other than fortuitous.  We see again that the physical significance of the BLM scale is questionable.

In summary, we have computed the $O(\beta_0 \alpha_s^2)$ corrections to top quark decay into W and Higgs bosons, and applied the BLM scale setting method.  For the case of decay into heavy bosons, we found extremely small BLM scales for processes that we intuitively expect to be calculable in perturbative QCD. The small BLM scales can be made larger by reexpressing decay rates as functions of the top quark running mass rather than the pole mass.  However, in this case the BLM scale exhibits unphysical dependence on the mass of the boson. These problems lead us to believe that for the BLM scale is not a useful tool for analyzing QCD corrections to these decays.  A more detailed probe of higher order corrections is needed in order to set the scale.

It is a pleasure to acknowledge helpful discussions with Adam Falk, George Chiladze, and Konstantin Matchev. This work was supported by the National Science Foundation under Grant No. PHY-9404057.


\begin{references}

\bibitem{CDF} F.~Abe {\it et al.}, Phys.\ Rev.\ {\bf D50} 2966(1994),
Phys.\ Rev.\ Lett.\ {\bf 74} 2626 (1995).

\bibitem{D0} S.~Abachi {\it et al.}, Phys.\ Rev.\ Lett.\ {\bf 74} 2632 (1995).

\bibitem{BLM} S.J.~Brodsky, G.P.~Lepage, and P.B.~Mackenzie, Phys.\ Rev.\ {\bf D28} 228 (1983).


\bibitem{REN} M.~Neubert, Phys.\ Rev.\ {\bf D51} 5924 (1995),CERN Preprint CERN
 -TH.7487/94 (1994), hep-ph 9412265; CERN Preprint CERN
 -TH.7524/94 (1995), hep-ph 9502264.

P.~Ball, M.~Beneke, and V.M.~Braun, Phys.\ Rev.\ {\bf D52} 3929 (1995),
     hep-ph 9503492.

\bibitem{BB} M.~Beneke, V.M.~Braun, Phys.\ Lett.\ {\bf B348} 513 (1995)

\bibitem{JK} M. Jezabek and J.H. Kuhn, Nucl.\ Phys.\ {\bf B314}, 1 (1989);
  Nucl.\ Phys.\ {\bf B320}, 20 (1989); Phys.\ Lett.\ {\bf B207}, 91 (1988).

C.S.~Li, R.~Oakes, and T.C.~Yuan, Phys.\ Rev.\ {\bf D43},3759 (1991).

A.~Czarnecki, Phys.\ Lett.\ {\bf B252} 467 (1990).


\bibitem{SV} B.H.~Smith and M.B.~Voloshin, Phys.\ Lett.\ {\bf B340} 176 (1994).

\bibitem{AC2} A.~Czarnecki, Acta Phys.~Pol.~B, 26 (1995), hep-ph 9503444.

\bibitem{LSW} M.~Luke, M.J.~Savage, and M.B.~Wise, Phys.\ Lett.\ {\bf B343} 329 (1995); Phys.\ Lett.\ {\bf B 345} 301 (1995).

\bibitem{LY}A.~Czarnecki and S.~Davidson, Phys.\ Rev.\ {\bf D48} 4183 (1993);  {\bf D47} 3063 (1993).
 
C.S.~Li and T.C.~Yuan, Phys.\ Rev.\ {\bf D42} 3088 (1990); {\bf D47} 2156(E) (1993).

J.~Liu and Y.P. Yao, Phys.\ Rev.\ {\bf D46} 4183 (1992).

\bibitem{BBZ} M.~Beneke, V.M.~Braun, and V.I.~Zakahrov, Phys.\ Rev.\ Lett.\ 73 (1994) 3058.
 
I.~Bigi {\it et al.}; Phys.\ Lett.\ 307 (1993) 154.

\bibitem{Mehen} T.~Mehen, work in progress.

\end{references}
\end{document}